\documentclass[prb,preprint]{revtex4-1}
\usepackage{amsmath}
\usepackage{amssymb}
\usepackage{graphicx}

\begin{document}
\title{Exploring the action landscape with trial world-lines}
\author{Yogesh N. Joglekar}
\email{yojoglek@iupui.edu}
\affiliation{Department of Physics, 
Indiana University Purdue University Indianapolis (IUPUI), 
Indianapolis, Indiana 46202, USA}
\author{Weng Kian Tham}
\affiliation{Department of Physics, 
Indiana University Purdue University Indianapolis (IUPUI), 
Indianapolis, Indiana 46202, USA}
\begin{abstract}
The Hamilton action principle, also known as the principle of least action, and Lagrange equations are an integral part of intermediate and advanced undergraduate mechanics. Although the Hamilton principle is oft stated as ``the action for any nearby trial world-line is greater than the action for the classical world-line", the landscape of action in the space of world-lines is rarely explored. Here, for three common problems from introductory physics - a free particle, a uniformly accelerating particle, and a simple harmonic oscillator - we present families of trial world-lines, characterized by a few  parameters, that evolve continuously from their respective classical world-lines. With explicit analytical expressions available for the action, they permit a graphical visualization of the action landscape in the space of nearby world-lines. Although these trial world-lines form only a subset of the space of all nearby world-lines, they provide a pedagogical tool that complements the traditional Lagrange equation approach and is well-suited for advanced undergraduate students.  
\end{abstract}
\maketitle

\section{Introduction}
\label{sec:intro}
Multivariate calculus and its applications to extremal problems are commonplace in undergraduate physics and mathematics.~\cite{kreyszig} A function $f(x_1,\ldots,x_n)$ of $n$ variables is stationary at a point if all partial derivatives of that function vanish at that point: $\partial f/\partial x_i=0$ for $i=1,\ldots,n$. Whether that point is a local maximum, minimum, or neither is determined by the symmetric second-derivative matrix $A_{ij}=\partial^2f/\partial x_i\partial x_j=A_{ji}$ at that point.~\cite{kreyszig} Thus, the problem of locating the stationary points of a function is equivalent to the problem of solving $n$ equations in $n$ variables, and can be tackled analytically or numerically.  A complementary approach for locating the stationary points is to look at the global landscape of the function $f(x_1,\ldots,x_n)$ in $\mathbb{R}^{n}$. This approach is typically emphasized~\cite{kreyszig} only for $n\leq 2$ when it is straightforward to obtain the graph ($n=1$) or the surface plot ($n=2$) of the function using ubiquitous software packages.~\cite{soft} Graphical exploration of the function in the vicinity of a stationary point also complements the standard technique used to determine whether the stationary point is a maximum, minimum, or a saddle point. However, this process becomes increasingly cumbersome when $n\geq 3$ and is therefore not emphasized in such cases. 

The Hamilton principle~\cite{goldstein,landau} (also known as the principle of least action~\cite{landau,feynman2}) is an integral part of advanced undergraduate physics. This principle states that for a particle in one dimension, among all world-lines $x(t)$ that connect two specified events $(x_1,t_1)$ and $(x_2,t_2)$,  the Hamilton action 
\begin{equation}
\label{eq:haction}
S[x(t)]=\int_{t_1}^{t_2} L(x(t),\dot{x}(t),t) dt
\end{equation}
is stationary for the classical world-line $x_c(t)$ that obeys Newton's second law.~\cite{goldstein,landau,feynman2} Here, $L(x,\dot{x},t)$ is the Lagrangian that, in general, depends upon the position $x(t)$ and velocity $\dot{x}=dx/dt$ of the particle, and may also explicitly depend upon time.  Since the action $S[x(t)]$ is now a {\it functional} of the world-line $x(t)$ - essentially a function of infinite variables $x(t)$ for $t_1< t< t_2$ - the criterion for the stationary world-line $x_c(t)$ is a differential equation, also known as Lagrange equation,
\begin{equation}
\label{eq:lagrange}
\left.\left(\frac{d}{dt}\frac{\partial L}{\partial\dot{x}}-\frac{\partial L}{\partial x}\right)\right|_{x=x_c}=0.
\end{equation}
The Hamilton action principle for particles and fields in arbitrary dimensions provides an intuitive way to emphasize the deep connections~\cite{feynman2,moore}  between diverse topics such as mechanics, electromagnetism, and quantum mechanics where the requirement of a stationary action leads to Newton's second law,~\cite{goldstein} Maxwell's equations,~\cite{goldstein} and the time-dependent Schr\"{o}dinger equation~\cite{merzbacher} respectively. There have been increasing efforts to introduce this topic in introductory physics courses.~\cite{moore,taylor,taylor2} These efforts, based on reorganizing the introductory mechanics content, are supplemented by interactive graphical software that ``hunts'' for the classical world-line through sampling.~\cite{software} Recently, the question of whether the action for the classical world-line is a minimum or a saddle point, and the fact that it is never a maximum, have been addressed in detail.~\cite{taylor} However, even in the simplest case of a particle in one dimension, complete numerical~\cite{software,bork} and graphical exploration of the global landscape of the action in the space of all world-lines is impossible. Therefore, although the Hamilton principle is oft stated~\cite{feynman2,class} as ``if you calculate the action for any world-line close to the classical world-line, it will be greater than the action for the classical world-line", this exercise is seldom carried out explicitly. 

In this article, we present families of trial world-lines characterized by a few parameters for a particle in constant, linear, and quadratic potentials. The classical motion of the particle in these cases is familiar to advanced physics undergraduate students. In each case, the trial world-lines in a family are continuously connected to the corresponding classical world-line. Using these trial world-lines, we graphically and analytically explore the action landscape in the vicinity of the classical world-line. In addition to explicitly verifying that the action is minimum for the classical world-line, this exercise permits a look at corrections to the action that arise when the trial world-lines deviate from the classical one. Due to the global nature of this approach, we are able to explore situations in which there are multiple classical world-lines that connect the two given events. We emphasize that by construction, the families of trial world-lines presented here are only a subset of the set of all nearby trial world-lines, and thus the action landscape obtained here is not exhaustive. This exercise provides, however, a way to explore the action landscape graphically using common software programs such as Mathematica, Grapher (Mac), or Sage.~\cite{sage} 

\section{The action landscape: from a functional to a function}
\label{sec:trial}
In this section we consider three common problems from introductory physics: a (almost) free particle, a uniformly accelerating particle, and a simple harmonic oscillator.~\cite{nr,introphysics} In each of these cases, the classical trajectory of the particle, typically obtained by using Newton's laws, is well-known to the students starting an intermediate or advanced mechanics course; therefore, in the following, we do not present the derivation of the classical world-lines.


\subsection{Free particle}
\label{subsec:fp}

\begin{figure}[h!]
\begin{center}
\begin{minipage}{20cm}
\begin{minipage}{9cm}
\hspace{-3cm}
\includegraphics[angle=0,width=8cm]{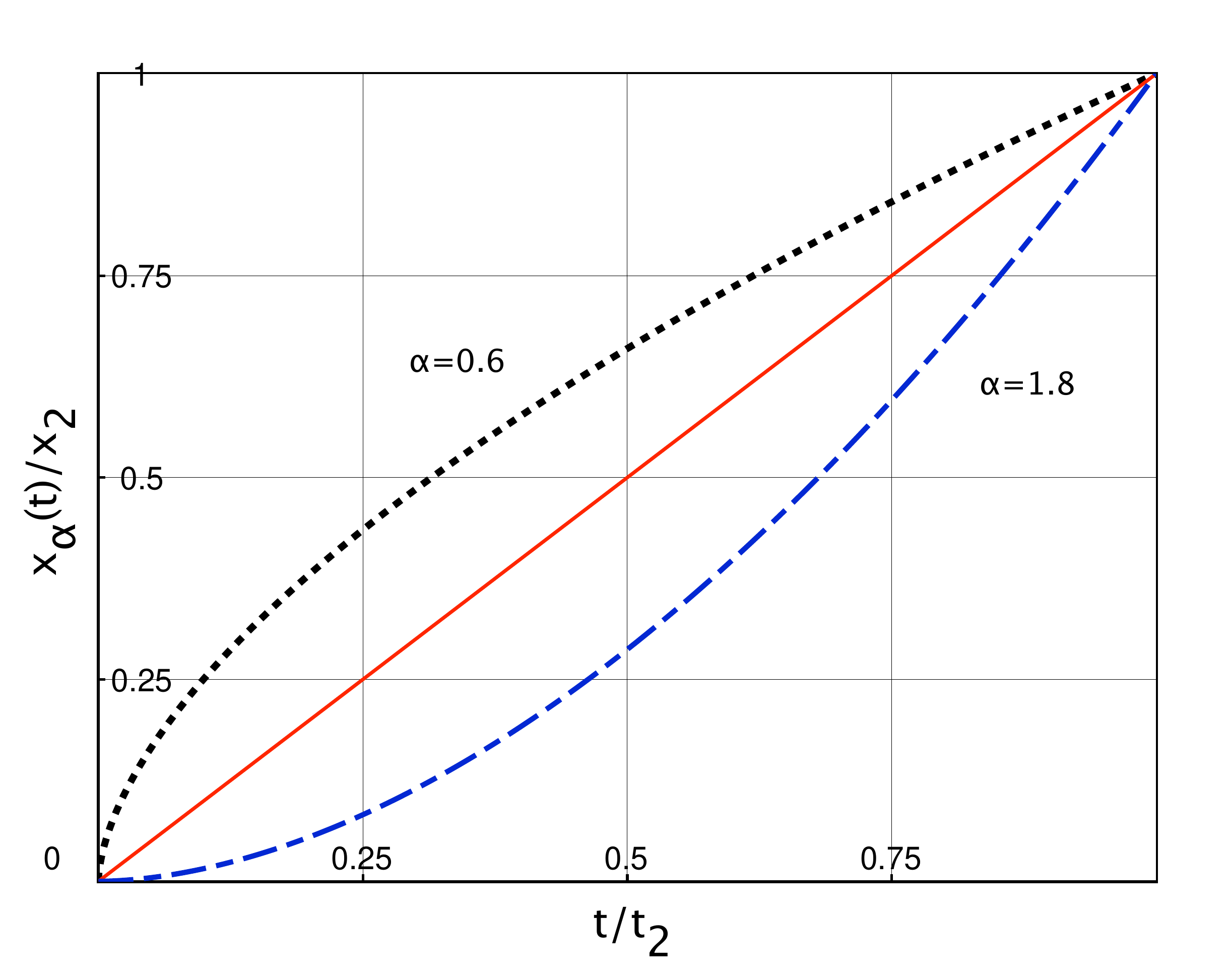}
\end{minipage}
\begin{minipage}{9cm}
\hspace{-5cm}
\includegraphics[angle=0,width=8cm]{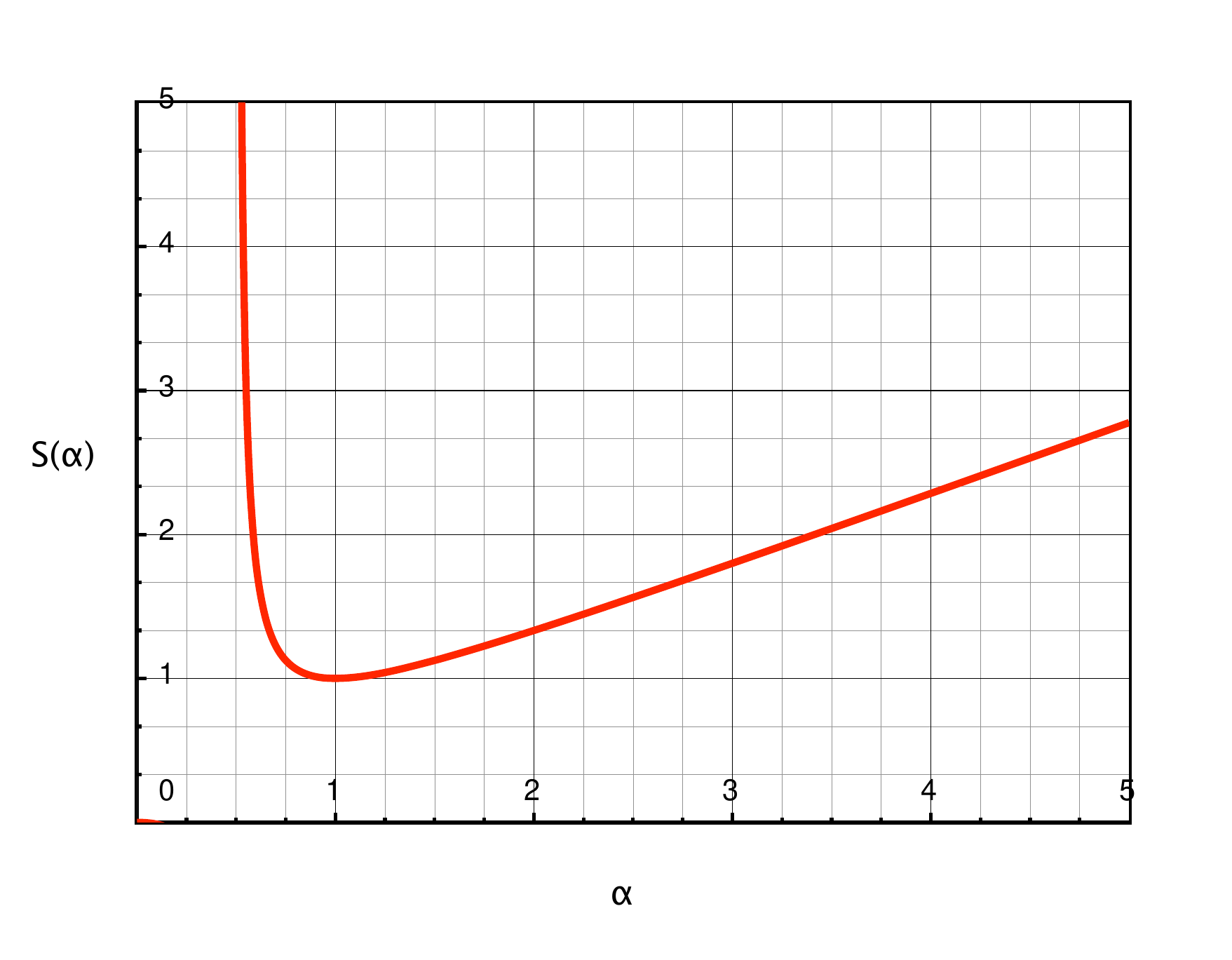}
\end{minipage}
\end{minipage}
\caption{\label{fig:alpha}
(color online) (a) The left panel shows trial world-lines for a free particle, $x_\alpha(t)=x_2(t/t_2)^\alpha$, for $\alpha$=0.6 (dotted line), $\alpha$=1.8 (dashed line), and the classical world-line $\alpha$=1 (solid line). Note that for trial world-lines with $\alpha<1$ the velocity diverges as $t\rightarrow 0$, for trial world-lines with $\alpha>1$ the velocity vanishes as $t\rightarrow 0$, and the action for the trial world-line is finite only when $\alpha>1/2$.  (b) The right panel shows that the dimensionless action $S(\alpha)/S_{FP}$ has the minimum at the classical world-line $\alpha$=1. As expected, the minimum action is equal to that for a free particle, $S_{FP}=mx_2^2/2t_2$, and has a quadratic expansion around the minimum, $\delta S=S_{FP}(\alpha-1)^2$.}
\end{center}
\end{figure}

Let us start with the case of a free particle of mass $m$ in one dimension. Since this system is translationally invariant, without loss of generality we choose $(x_1,t_1)=(0,0)$. The classical world-line connecting the initial event $(0,0)$ and the final event $(x_2,t_2)$ is given by $x_c(t)=x_2 t/t_2$, and, since the Lagrangian for this system is $L(x,\dot{x})=m\dot{x}^2/2$, the corresponding free-particle action is given by $S_{FP}=m x_2^2 /2t_2$. Let us consider a family of trial world-lines $x_\alpha(t)=x_2 (t/t_2)^\alpha$ characterized by a single parameter $\alpha$. It is straightforward to evaluate the action as a function of $\alpha$
\begin{equation}
\label{eq:alpha}
S(\alpha)= S_{FP}\left[\frac{\alpha^2}{2\alpha-1}\right].
\end{equation}
Note that a well-defined action requires $\alpha>1/2$. The action is minimum when $dS/d\alpha=S_{FP}2\alpha(\alpha-1)/(2\alpha-1)^2=0$ or, equivalently, $\alpha=1$. Fig.~\ref{fig:alpha} shows typical trial world-lines along with the classical world-line at $\alpha=1$ (solid line), and dimensionless action $S(\alpha)/S_{FP}$ as a function of the exponent $\alpha$. We see that the graph has a single minimum at $\alpha=1$. A Taylor expansion of the action (\ref{eq:alpha}) shows that for nearby world-lines, $\alpha\approx 1$, the change in the action $\delta S=S(\alpha)-S_{FP}$ varies quadratically with the distance from the classical world-line, $\delta S=S_{FP}(\alpha-1)^2$, although, as we will see below, this conclusion is not generic.

\begin{figure}[h!]
\begin{center}
\begin{minipage}{20cm}
\begin{minipage}{9cm}
\hspace{-3cm}
\includegraphics[angle=0,width=8cm]{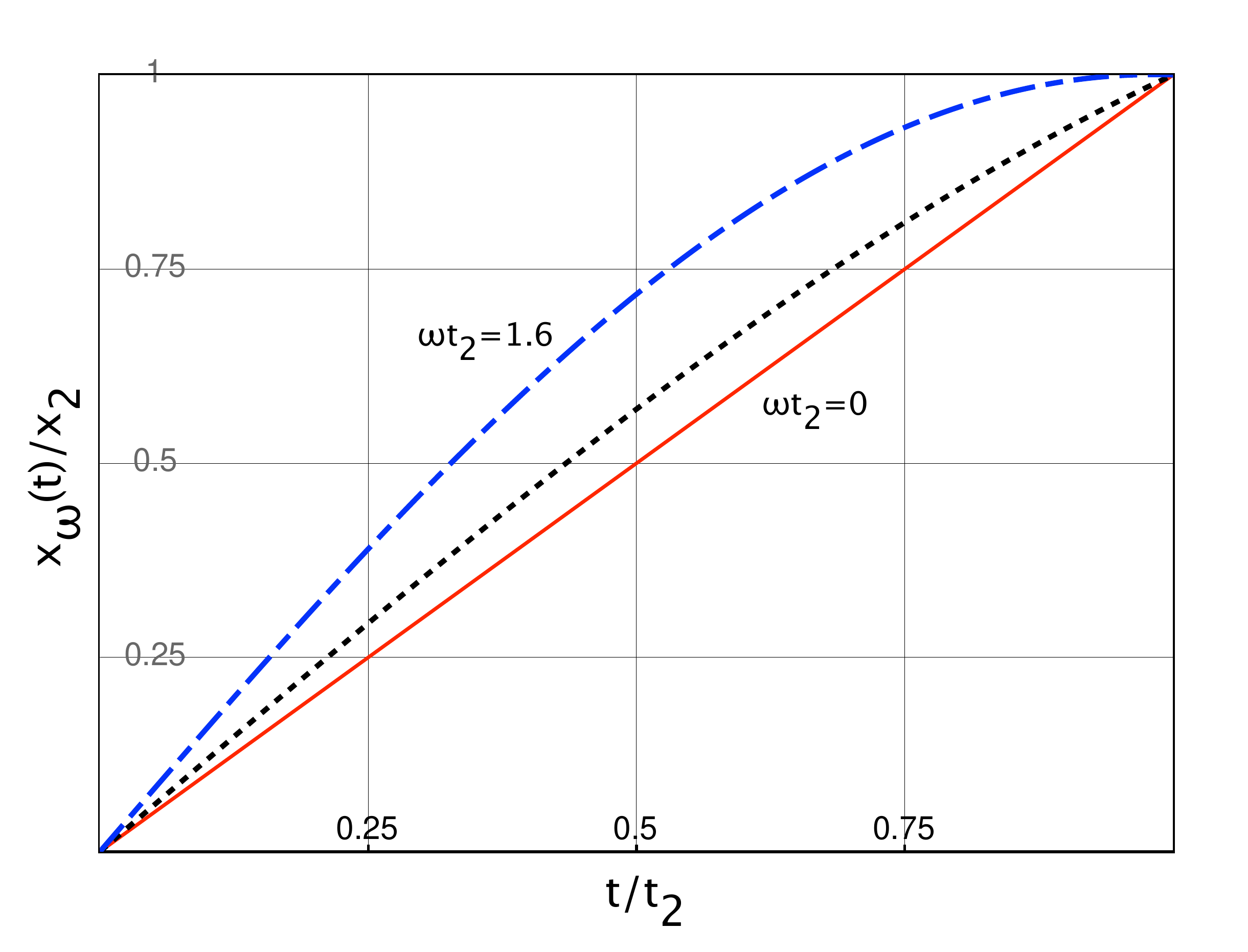}
\end{minipage}
\begin{minipage}{9cm}
\hspace{-5cm}
\includegraphics[angle=0,width=8cm]{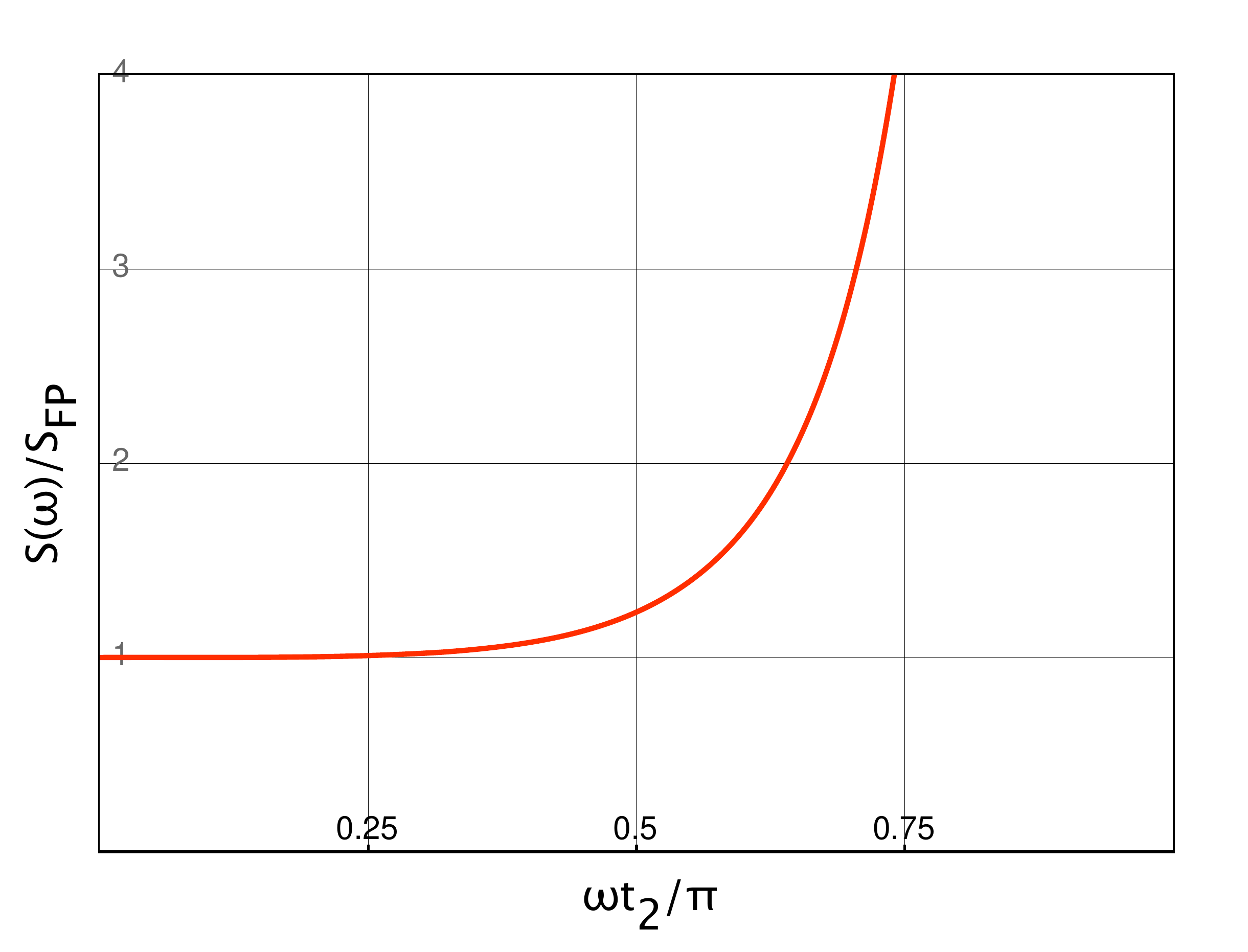}
\end{minipage}
\end{minipage}
\caption{\label{fig:omega}
(color online)  (a) The left panel shows trial world-lines for a free particle, $x_\omega(t)=x_2\sin\omega t/\sin\omega t_2$ for $\omega t_2$=1.6 (dashed line), $\omega t_2$=0.8 (dotted line), and the classical world-line $\omega t_2$=0 (solid line). Note that $x_\omega(t)$ is an even function of $\omega$. Since the global minimum of Eq.(\ref{eq:omega}) occurs in the range $-\pi<\omega t_2<\pi$, we only focus 
on $|\omega t_2|<\pi$. (b) The right panel shows the dimensionless action $S(\omega)/S_{FP}$ restricted to $\omega\geq 0$ since $S(\omega)$ is an even function of $\omega$.  The action shows the minimum at the classical world-line $\omega$=0, and the {\it flatness} of the curve near $\omega=0$ is consistent with the quartic expansion, $\delta S(\omega)=S_{FP}(\omega t_2)^4/45$.}
\end{center}
\end{figure}

Another family of trial world-lines for the same problem is given by $x_\omega(t)= x_2\sin(\omega t)/\sin(\omega t_2)$ for $\omega\neq 0$ and $x_{\omega=0}(t) = x_2 t/t_2$. Note that $x_\omega(t)$ is an even, continuous function of $\omega$. The Hamilton action for the trial world-line is given by 
\begin{equation}
\label{eq:omega}
S(\omega)=S_{FP}\frac{(\omega t_2)^2}{2\sin^2(\omega t_2)}\left[1 +\frac{\sin(2\omega t_2)}{(2\omega t_2)}\right]=S(-\omega).
\end{equation}
The trial action diverges at $\omega t_2=n\pi$, and has a single minimum at $n\pi<\omega t_2<(n+1)\pi$ between consecutive divergences. It follows from Eq.(\ref{eq:omega}) that the global minimum occurs when $\omega t_2\in(-\pi,\pi)$. Since the action is an even function of the parameter $\omega$, it follows that it is either a minimum or a maximum, but not a saddle point, at $\omega=0$, the classical world-line. Taylor expansion of Eq.(\ref{eq:omega}) around $\omega=0$ shows that the change in the action $\delta S$ for this family of trial world-lines is quartic in the distance from the classical world-line, $\delta S= +S_{FP} (\omega t_2)^4/ 45$. Figure~\ref{fig:omega} shows typical trial world-lines, along with the classical world-line $\omega=0$ (solid line), and the dimensionless action $S(\omega)/S_{FP}$ as a function of the dimensionless parameter $\omega t_2$. We see that, indeed, the action $S(\omega)$ has a {\it very flat} minimum at $\omega=0$. These two families of trial world-lines show that even for a free particle, the action landscape in the vicinity of the classical world-line is very rich.  

\subsection{Free particle with a perfectly reflecting barrier}
\label{subsec:barrier}
\begin{figure}[h!]
\begin{center}
\includegraphics[angle=0,width=12cm]{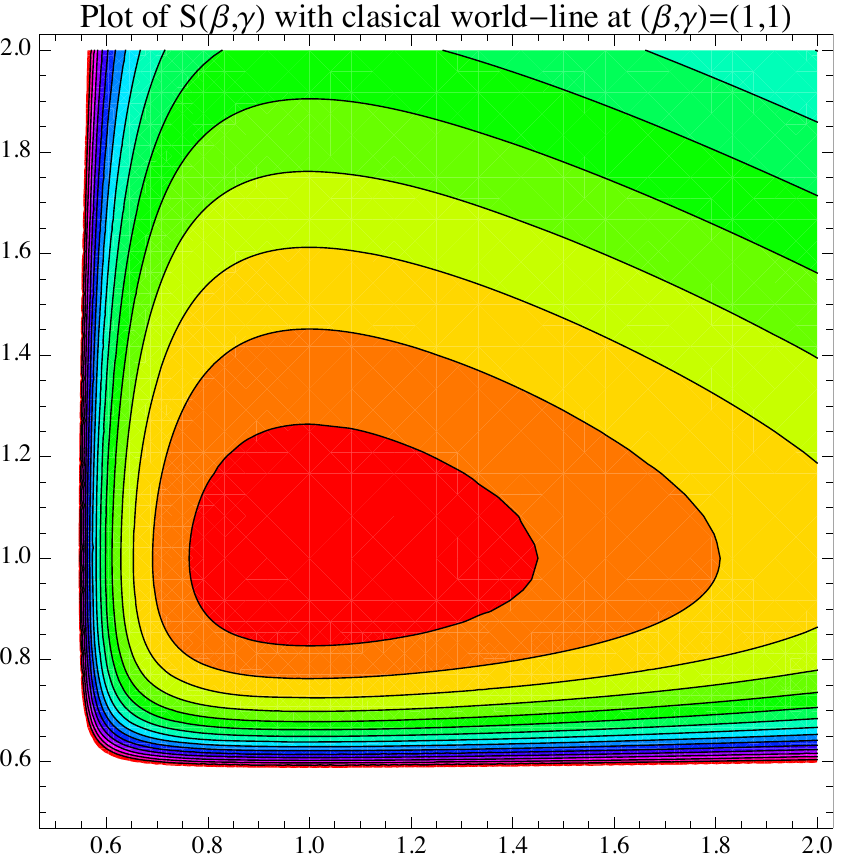}
\caption{\label{fig:actionbetagamma}
(color online) Typical contour plot of the action $S(\beta,\gamma)/S_2$, Eq.(\ref{eq:actionbetagamma}), with trial world-lines $x_{\beta,\gamma}(t)$ for a free particle bouncing off a barrier located at $X_b=-3x_2/4$ or, equivalently, $\xi=0.3$. The classical world-line $x_{c2}(t)$, Eq.(\ref{eq:xc2}), corresponds to $(\beta,\gamma)=(1,1)$ where the action shows a clear minimum. Note that as $\xi\rightarrow 1/2$, the action anisotropy in the vicinity of the classical world-line in the $\beta$-$\gamma$ plane vanishes.} 
\end{center}
\end{figure}

Now, we consider the free particle with a world-line that starts at $(x_1,t_1)=(0,0)$ and ends to the right at $x_2>0$ at time $t_2>0$, in the presence of a perfectly reflecting barrier at $X_b\leq 0$. In this case, the particle is free at all times except when it collides with the barrier; at that instance, its momentum is reversed while its kinetic energy is unchanged. This system provides the simplest example where there are two classical world-lines that satisfy the given boundary conditions, and without further information (say, about the direction of the initial velocity) it is impossible to choose a unique classical world-line. It is useful, for the purposes of analytical calculations, to define the barrier position as $X_b=-\eta x_2$ where $\eta\geq 0$ is a dimensionless scale factor. Due to the presence of the barrier, in addition to the classical world-line $x_{c1}(t)=x_2 t/t_2$, there is another world-line that represents the particle initially moving to the left, bouncing off the barrier, and then traveling to the right, 
\begin{equation}
\label{eq:xc2}
x_{c2}(t)=\left\{
\begin{array}{cc}
-\eta x_2 t/(\xi t_2) &  0\leq t\leq \xi t_2,\\
-\eta x_2+ (1+\eta)x_2(t-\xi t_2)/\left[(1-\xi)t_2\right] &   \xi t_2\leq t\leq t_2.\\
\end{array}\right.
\end{equation}
Here $\xi=\eta/(2\eta+1)< 1/2$ represents the fraction of the time that the particle spends moving to the left. Note that the speed of the particle along the second classical world-line $x_{c2}(t)$ is $(2\eta+1)$ times larger than its speed along the first classical world-line $x_{c1}(t)$. Hence the actions for the two classical world-lines are given by $S_{1}=mx_2^2/2t_2=S_{FP}$ and $S_2=(2\eta+1)^2S_1$ respectively. Since we have already studied the action landscape around the first world-line, we will focus on the second. To explore the action landscape in the vicinity of the second classical world-line, we consider a family of trial world-lines that are parametrized by two exponents
\begin{equation}
\label{eq:xc2betagamma}
x_{\beta,\gamma}(t)=\left\{
\begin{array}{cc}
-\eta x_2 \left[t/(\xi t_2)\right]^\beta &  0\leq t\leq \xi t_2,\\
-\eta x_2+ (1+\eta)x_2\left[(t-\xi t_2)/(1-\xi)t_2\right]^\gamma &   \xi t_2\leq t\leq t_2.\\
\end{array}\right.
\end{equation}
Note that, as before, a well-defined action requires $\beta,\gamma >1/2$, and that the second classical world-line corresponds to $(\beta,\gamma)=(1,1)$. The Hamilton's action for the trial world-line is given by 
\begin{equation}
\label{eq:actionbetagamma}
S(\beta,\gamma)= S_2\left[\xi\frac{\beta^2}{(2\beta-1)}+(1-\xi)\frac{\gamma^2}{(2\gamma-1)}\right].
\end{equation}
This action is a sum of actions for the two paths characterized by $\beta$ and $\gamma$ weighted by the fraction of the time that the particle spends on each path. Since the trial action is now a function of two variables, it is stationary when $\partial S/\partial\beta=0=\partial S/\partial\gamma$. It is straightforward to verify that $(\beta,\gamma)=(1,1)$, which corresponds to the second classical world-line $x_{c2}(t)$, is the only solution.  Since the time that the particle spends moving left cannot be greater than the time it spends moving right, $\xi< 1/2< (1-\xi)$, the action $S(\beta,\gamma)$ is anisotropic around $(\beta,\gamma)=(1,1)$, and the anisotropy vanishes only when $\xi\rightarrow 1/2$. Figure~\ref{fig:actionbetagamma} shows a typical contour plot of the dimensionless action $S(\beta,\gamma)/S_2$ in the $(\beta,\gamma)$ plane for a particle with barrier at $X_b=-3x_2/4$ or, equivalently, $\xi=0.3$. As expected, we see that the action is minimum at $(\beta,\gamma)=(1,1)$. 

Next, we can consider a set of trial world-lines that span {\it both} classical world-lines: $x_{p,\alpha,\beta,\gamma}(t) = p x_\alpha(t)+ (1-p) x_{\beta,\gamma}(t)$. Note that the first classical world-line $x_{c1}(t)$ is obtained when $(p,\alpha)=(1,1)$ whereas the second classical world-line $x_{c2}(t)$  is obtained when $(p,\beta,\gamma)=(0,1,1)$. We leave it as an exercise for the Reader to obtain the action $S(p,\alpha,\beta,\gamma)$: in order to simplify the calculation, choose $\alpha=\beta=\gamma$; plot the dimensionless action $S(p,\alpha)/S_{FP}$ in the $(p,\alpha)$ plane and verify that the action has local minima at the two classical world-lines.    


\subsection{Uniformly accelerating particle}
\label{subsec:force}
\begin{figure}[h!]
\begin{center}
\begin{minipage}{20cm}
\begin{minipage}{9cm}
\hspace{-3cm}
\includegraphics[angle=0,width=8cm]{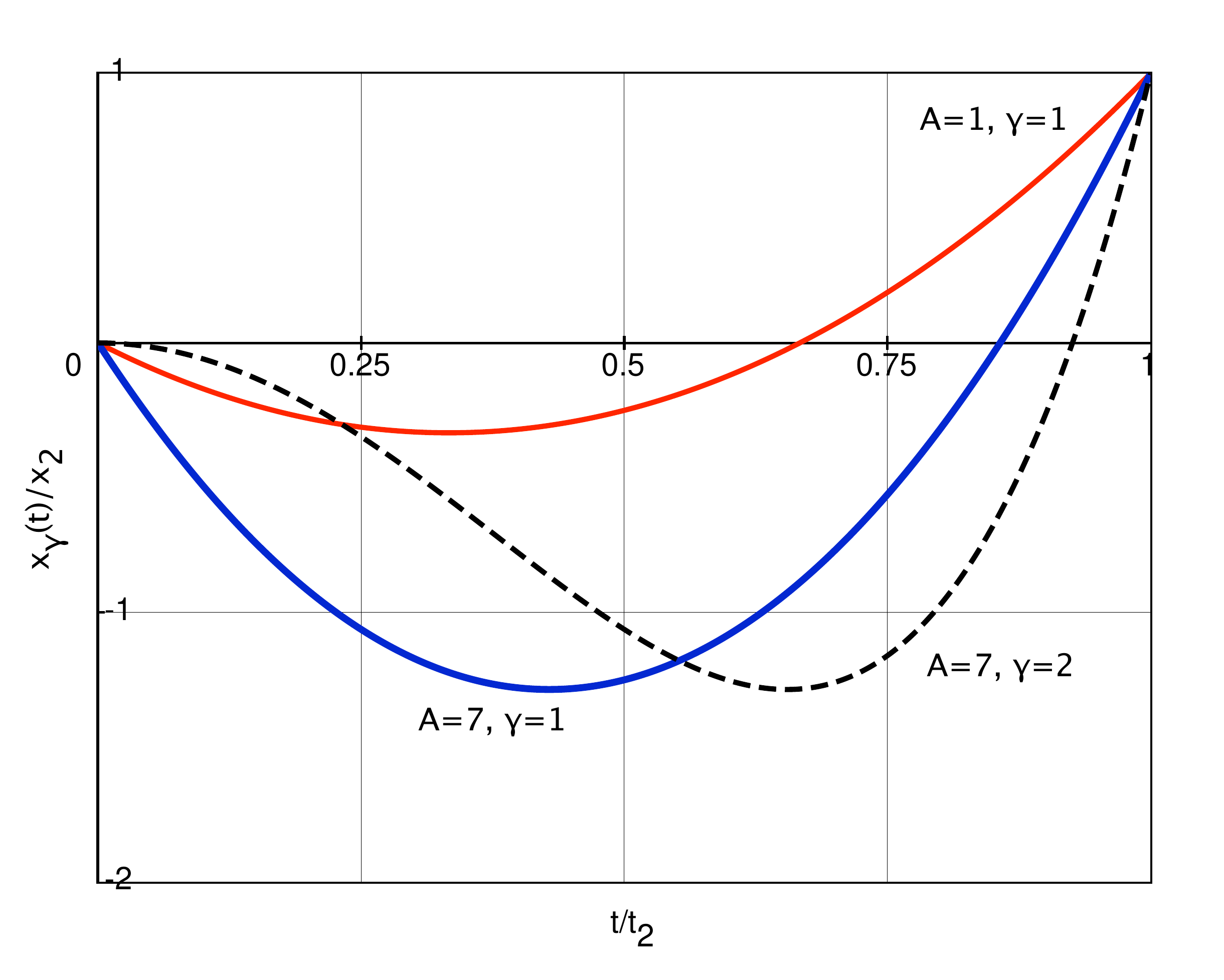}
\end{minipage}
\begin{minipage}{9cm}
\hspace{-5cm}
\includegraphics[angle=0,width=8cm]{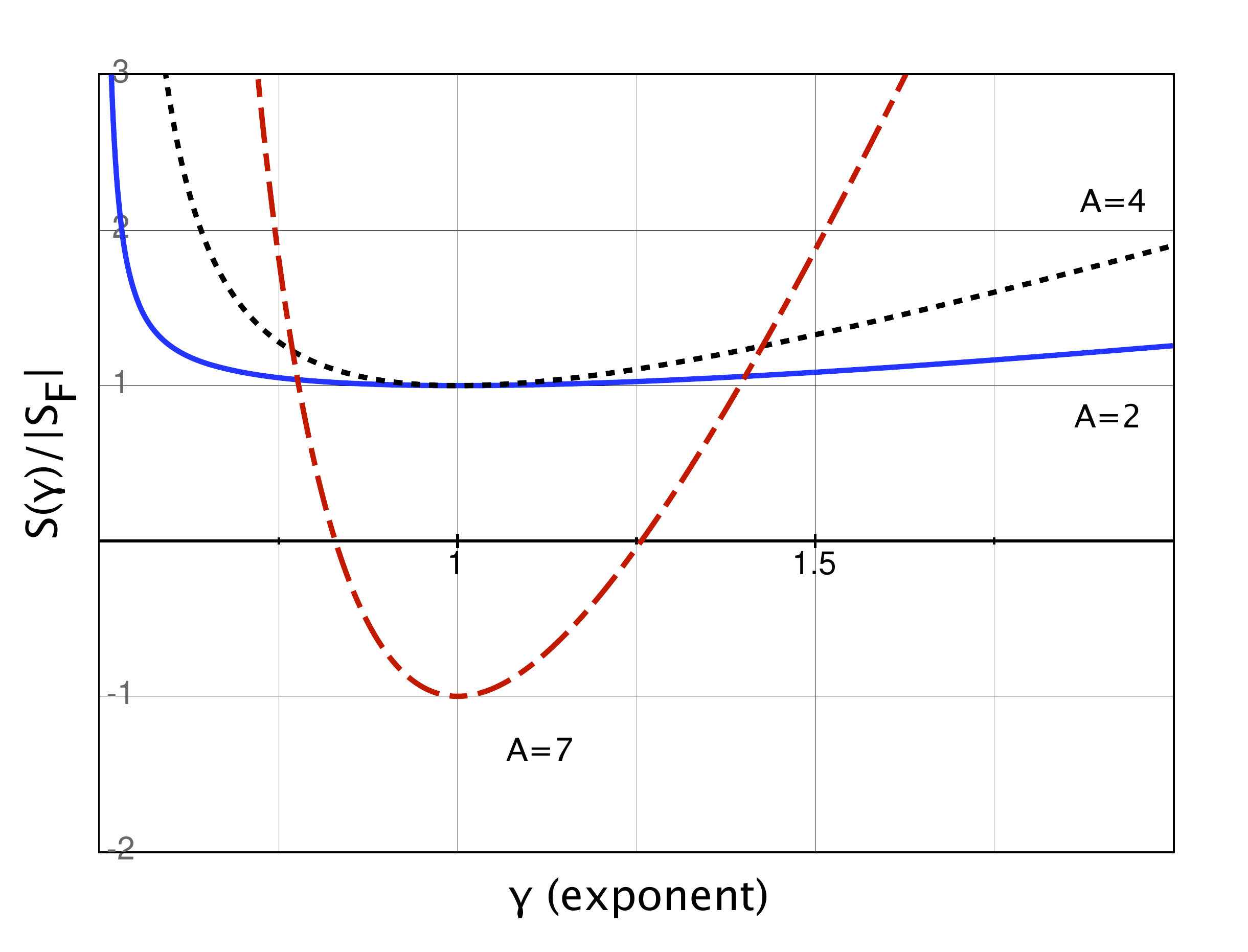}
\end{minipage}
\end{minipage}
\caption{\label{fig:gamma}
(color online)(a) The left panel shows trial world-lines for a uniformly accelerated  particle
$x_\gamma(t)=x_2\left[(1-A)u^\gamma + Au^{2\gamma}\right]$ where $\gamma=1$ corresponds to the classical world-line for any $A=Ft_2^2/2mx_2$. Note that as $A$ increases the time the particle spends in the region $x<0$ increases; hence the action for the trial path decreases and becomes negative when $A>A_c=(3+2\sqrt{3})$. (b) The right panel shows the dimensionless action as a function of the trial world-line parameter $\gamma$ for various values of $A$. For $A<A_c$ (dotted line and solid line), we see that the action has a minimum value of +1 at $\gamma=1$, whereas for $A>A_c$ (dashed line), the action has a minimum value of $-1$ at $\gamma=1$. In each case, the action increases quadratically as $\gamma$ deviates from the classical world-line value $\gamma=1$.}
\end{center}
\end{figure}

We now consider a particle of mass $m$ in one-dimension being acted upon by a constant force $F$. Since this system, too, is translationally invariant, we choose $(x_1,t_1)=(0,0)$. The classical world-line joining the two events is given by 
\begin{equation}
\label{eq:Fpath}
x_c(t)= \left(x_2-\frac{F t_2^2}{2m}\right)\left(\frac{t}{t_2}\right)+\left(\frac{F t_2^2}{2m}\right)\left(\frac{t}{t_2}\right)^2=x_2\left[(1-A)u +A u^2\right],
\end{equation}
where $A=F t_2^2/2mx_2$ is the dimensionless acceleration and $u=t/t_2\in [0,1]$ is the dimensionless time variable. Since the Lagrangian for this system is $L(x,\dot{x})= m\dot{x}^2/2 + Fx$, the action for the classical world-line is given by $S_{F}=(mx_2^2/2t_2)[1+2A - A^2/3]= S_{FP}[1+2A - A^2/3]$. It increases continuously from its free-particle value when $0\leq A<3$, and is monotonically decreasing for $A>3$ becoming negative when $A\ge A_c=(3+2\sqrt{3})$. Motivated by Eq.(\ref{eq:Fpath}), we consider a family of trial world-lines characterized by a single parameter $\gamma$ and the corresponding action,
\begin{eqnarray}
\label{eq:gamma}
x_{\gamma}(t) & = & x_2\left[(1-A)u^\gamma+A u^{2\gamma}\right],\\
\label{eq:actiongamma}
S(\gamma) & = & S_{FP}\left[\frac{(1-A)^2\gamma^2}{(2\gamma-1)}+\frac{4A^2\gamma^2}{(4\gamma-1)}+\frac{4A(1-A)\gamma^2}{(3\gamma-1)}+\frac{4A^2}{(2\gamma+1)}+\frac{4A(1-A)}{(\gamma+1)}\right].
\end{eqnarray}
Note that once again, a well-defined action requires $\gamma>1/2$. The action is stationay for any value of $A$ when $dS/d\gamma=0$ or equivalently when $\gamma=1$, the classical world-line. A Taylor expansion of Eq.(\ref{eq:actiongamma}) near $\gamma=1$ shows that for nearby trial world-lines, the change in the action is  positive, $\delta S = +S_{FP}(20 A^2/27-A+1)(\gamma-1)^2>0$ for any $A$, and thus confirms that $\gamma=1$ is a minimum. Figure~\ref{fig:gamma} shows the trial world-lines for $A=1$ and $A=7$ including the classical world-lines (solid lines), and the dimensionless action $S(\gamma)/|S_F|$ as a function of $\gamma$ for different values of $A$. The action has a single minimum at $\gamma=1$, and the minimum value changes from +1 to  $-1$ when $A>A_c$ as is expected. Thus, Eq.(\ref{eq:gamma}) provides a suitable one-parameter family of trial world-lines to explore the action landscape for a uniformly accelerated particle.

\begin{figure}[h!]
\begin{center}
\includegraphics[angle=0,width=12cm]{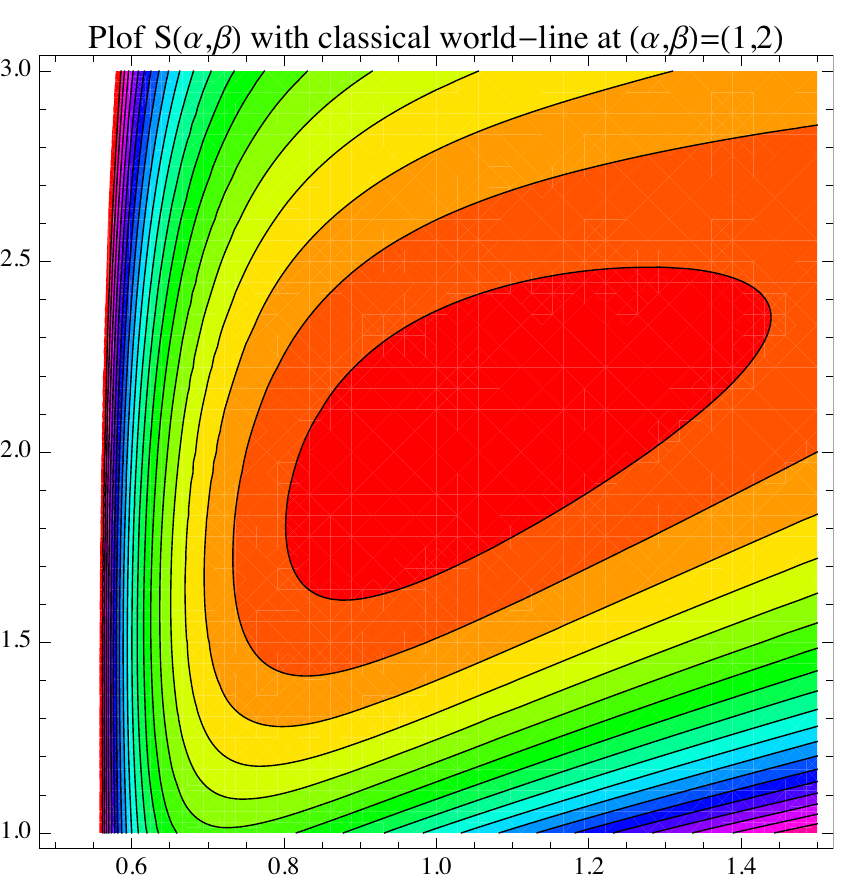}
\caption{\label{fig:actionalphabeta}
(color online) Typical contour plot of the action $S(\alpha,\beta)/|S_{FP}|$, Eq.(\ref{eq:actionalphabeta}), with trial world-lines $x_{\alpha\beta}(t)$ for a uniformly accelerated particle with $A=3<A_c$. The action shows a clear minimum at the classical world-line $(\alpha,\beta)=(1,2)$. We note that as the value of $A$ is increased, the anisotrophy in the $\alpha$-$\beta$ plane in the vicinity of the classical world-line increases.}
\end{center}
\end{figure}

For the same system, an alternate family of trial world-lines parameterized by two exponents and the corresponding action are given by
\begin{eqnarray}
\label{eq:alphabeta}
x_{\alpha,\beta}(t) & = &  x_2\left[(1-A)u^\alpha+ A u^\beta\right], \\
\label{eq:actionalphabeta}
S(\alpha,\beta) & = & S_{FP}\left[\frac{(1-A)^2\alpha^2}{(2\alpha-1)}+\frac{A^2\beta^2}{(2\beta-1)}+\frac{2A(1-A)\alpha\beta}{(\alpha+\beta-1)}+\frac{4A^2}{(\beta+1)}+\frac{4A(1-A)}{(\alpha+1)}\right].
\end{eqnarray}
Note that, as expected, Eq.(\ref{eq:actionalphabeta}) reduces to Eq.(\ref{eq:actiongamma}) when
$\beta=2\alpha$ and that a well-defined action requires $\alpha,\beta>1/2$. The trial world-line $x_{\alpha,\beta}(t)$ is continuously connected to the classical world-line, Eq.(\ref{eq:Fpath}), that occurs when $(\alpha,\beta)=(1,2)$. The action is stationary when $\partial S/\partial\alpha=0=\partial S/\partial\beta$. It is possible, albeit tedious, to analytically verify that $(\alpha,\beta)=(1,2)$ is a solution of these equations, and that the matrix of second derivatives at this point has positive eigenvalues confirming that it is a minimum. Figure~\ref{fig:actionalphabeta} shows a typical contour plot of the dimensionless action $S(\alpha,\beta)/|S_F|$ in the $\alpha$-$\beta$ plane for $A=3<A_c$. The action shows a clear minimum at the classical world-line $(\alpha,\beta)=(1,2)$. We also note that action landscape in the vicinity of the classical world-line becomes increasingly anisotropic as $A$ increases. Thus Eqs.(\ref{eq:gamma}) and (\ref{eq:alphabeta}) provide two distinct ways to graphically explore the action landscape in the vicinity of the classical world-line, Eq.(\ref{eq:Fpath}).  


\subsection{Simple harmonic oscillator}
\label{subsec:sho}
\begin{figure}[h!]
\begin{center}
\includegraphics[angle=0,width=12cm]{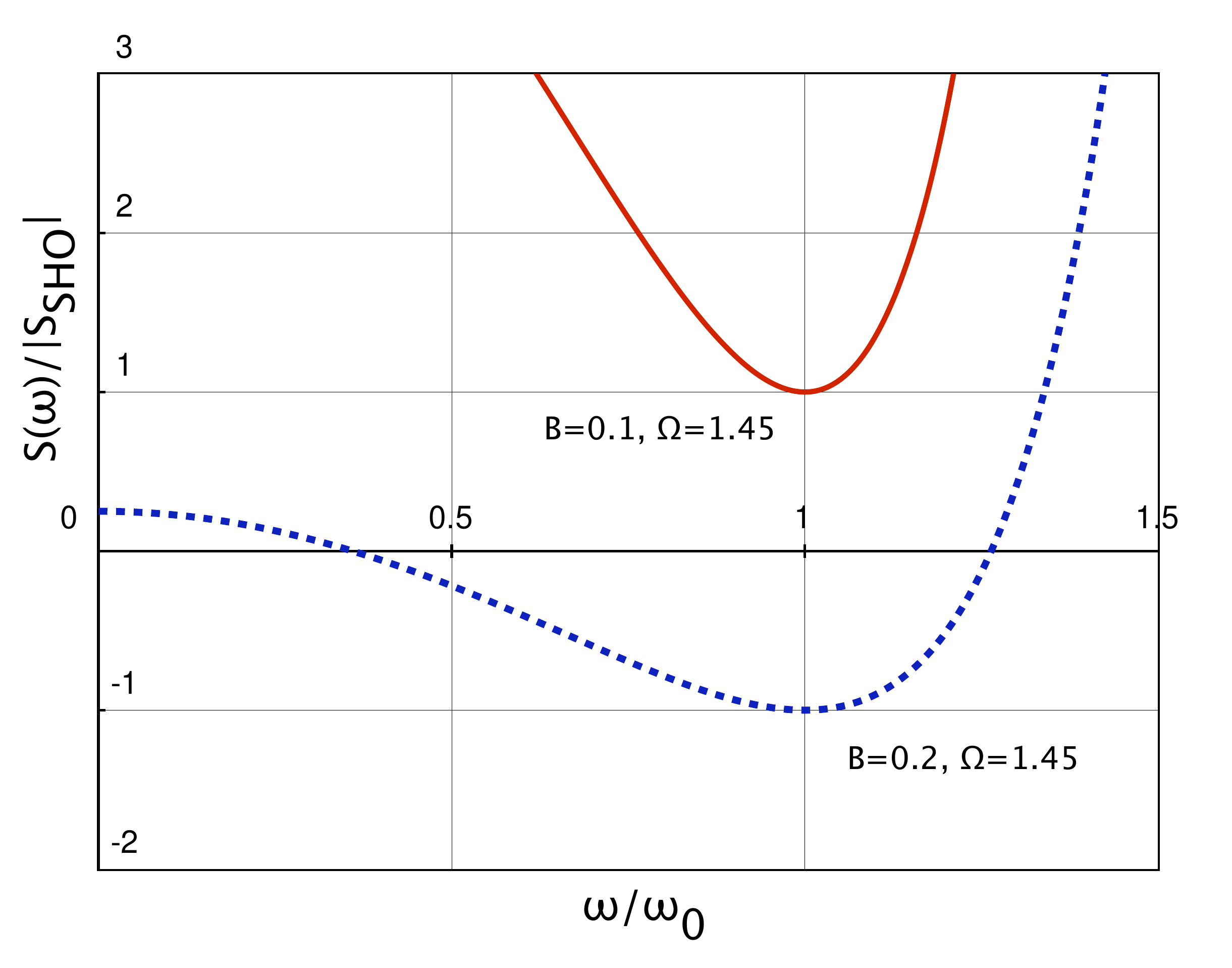}
\caption{\label{fig:sho}
(color online) Typical plots of the action $S(\omega)$ for a simple harmonic oscillator for trial world-lines $x_\omega(t)$, Eq(\ref{eq:shoomega}). The action is plotted in units of absolute value of the classical action $S_{SHO}$ and the frequency $\omega$ is in units of the natural frequency $\omega_0$ of the system. For all values of $B=2x_1x_2/(x_1^2+x_2^2)$ and $\Omega=\omega_0\Delta t$, the action $S(\omega)$ shows the minimum at $\omega=\omega_0$. For a fixed $\Omega=1.45$ as $B$ is increased from $B=0.1$ (solid line) to $B=0.2$ (dotted line) , $\Omega_c=\arccos(B)$ is reduced, and hence the action minimum changes its value from +1 (solid line) to $-1$ (dotted line).}
\end{center}
\end{figure}

As the last example, let us consider the particle of mass $m$ in a simple harmonic potential given by $V(x)=m\omega_0^2x^2/2$. The classical world-line $x_c(t)$ connecting events $(x_1,t_1)$ and $(x_2,t_2)$ is given by~\cite{hibbs,wen}
\begin{equation}
\label{eq:shoclassical}
x_c(t)  =  x_2\frac{\sin\omega_0(t-t_1)}{\sin\omega_0(t_2-t_1)}+x_1\frac{\sin\omega_0(t_2-t)}{\sin\omega_0(t_2-t_1)}.
\end{equation}
Note that Eq.(\ref{eq:shoclassical}) is invalid when $\omega_0(t_2-t_1)= n\pi$ or, equivalently when $x_2=\pm x_1$. In these two cases the action for the classical world-line is zero and thus it is impossible to quantify ``a small change to the classical action" without resorting to the fundamental (quantum) unit of action $\hbar$.  Since the Lagrangian for this system is given by $L(x,\dot{x})=m\dot{x}^2/2-m\omega_0^2x^2/2$, it is straightforward to evaluate the action for the classical world-line and we get~\cite{hibbs,wen}
\begin{eqnarray}
S_{SHO} & = & m\omega_0\left[(x_1^2+x_2^2)\cos\omega_0(t_2-t_1)-2x_1x_2\right]/2\sin\omega_0(t_2-t_1),\\
& = & m(x_1^2+x_2^2)\omega_0^2\Delta t\left[\frac{\cos\Omega-B}{2\Omega\sin\Omega}\right],
\end{eqnarray}
where $\Delta t=(t_2-t_1)$, and $\Omega=\omega_0\Delta t\neq n\pi$ and $B=2x_1x_2/(x_1^2+x_2^2)$ are dimensionless quantities. Since $|B|<1$ the classical action changes from positive to negative when $\Omega>\Omega_c=\arccos(B)$.  Note that Eq.(\ref{eq:shoclassical}) satisfies the classical world-line boundary conditions for any $\omega$. Therefore we consider a one-parameter family of trial world-lines
\begin{equation}
\label{eq:shoomega}
x_\omega(t)=  x_2\frac{\sin\omega(t-t_1)}{\sin\omega(t_2-t_1)}+x_1\frac{\sin\omega(t_2-t)}{\sin\omega(t_2-t_1)}.
\end{equation}
The corresponding Hamilton action is given by 
\begin{equation}
S(u)=\frac{m(x_1^2+x_2^2)\omega_0^2\Delta t}{4\sin^2\Omega u}\left\{(u^2-1)\left(1-B\cos\Omega u\right)+(u^2+1)\left[\frac{\sin 2\Omega u}{2\Omega u}-B\frac{\sin\Omega u}{\Omega u}\right]\right\},
\end{equation}
where $u=\omega/\omega_0$ is the dimensionless parameter. It is straightforward to analytically check that the action has a minimum at $u=1$, $dS/du|_{u=1}=0$ and $d^2S/du^2|_{u=1}>0$. Figure~\ref{fig:sho} shows the typical dimensionless action $S(u)/|S_{SHO}|$ for $\Omega=1.45$ when $B=0.1$ (solid line) and $B=0.2$ (dotted line).  In each case, the action has the minimum at $u=1$. We recall that as $B$ increases, $\Omega_c=\arccos(B)$ decreases; therefore, for a fixed value of $\Omega$, the action minimum changes from +1 at $B=0.1$ (solid line) to $-1$ at $B=0.2$ (dotted line). We see from the graph that the action for a trial world-line deviates quadratically with the distance from the classical world-line, $\delta S\propto (u-1)^2$; in addition, its curvature at $u=1$ diverges as $\Omega\rightarrow\Omega_c$ from both above and below. It is straightforward to extend this analysis to a family of trial world-lines characterized by two parameters,
\begin{equation}
\label{eq:munu}
x_{\mu,\nu}(t)=x_2\frac{\sin\mu(t-t_1)}{\sin\mu(t_2-t_1)}+x_1\frac{\sin\nu(t_2-t)}{\sin\nu(t_2-t_1)}.
\end{equation}
In this family, the classical world-line occurs when $(\mu,\nu)=(\omega_0,\omega_0)$. We leave it as an exercise for the Reader to obtain an analytical expression for the action $S(\mu,\nu)$ and to graphically explore the dimensionless action $S(\mu,\nu)/|S_{SHO}|$ in the $(\mu,\nu)$ plane in the vicinity of the classical world-line.

\section{Conclusions}
\label{sec:conclusions}
The principle of least action and Lagrange equations are two integral elements of intermediate and advanced undergraduate mechanics. Subsequently, they are taught in advanced electromagnetism and quantum mechanics courses to emphasize the deep connections among the aforementioned subfields of physics. In all instances, however, an explicit calculation of the Hamilton's action for ``nearby trial paths'' is seldom carried out. In this article, we have graphically and analytically explored the action landscape in the vicinity of the classical world-line. Since the action is a {\it functional} of a given world-line $x(t)$ that obeys the boundary conditions, it is impossible to exhaustively span all trial world-lines close to the true classical world-line $x_c(t)$. Instead, for three familiar problems from introductory physics, we have presented families of trial world-lines that are characterized by a few parameters and that evolve continuously from their respective classical world-lines. Within each of these families, the Hamilton's action is reduced from a {\it functional} to a function of the (few) parameters that characterize the family. The calculation of action for these trial world-lines is accessible to advanced undergraduate students. Its graphical exploration using common software~\cite{soft,sage} permits an explicit, albeit limited, demonstration of the principle of least action with an approach that is complementary to the Lagrange equations.~\cite{goldstein,landau,feynman2} 

\begin{acknowledgments}
We wish to thank the anonymous referee for constructive criticism and useful suggestions.
\end{acknowledgments}


\section*{Figure captions}

\begin{thebibliography}{99}
\bibitem{kreyszig} E. Kreyszig, {\it Advanced Engineering Mathematics} (Wiley, New York, 2010).
\bibitem{soft} See, for example, Mathematica, Grapher (Mac), Sage,~\cite{sage} and numerous apps available at the Apple App Store.
\bibitem{goldstein} H. Goldstein, C.P. Poole, and J.L. Safko, {\it Classical Mechanics} (Addison-Wesley, San Francisco, 2002), pp. 34-35.
\bibitem{landau} L.D. Landau and E.M. Lifshitz, {\it Mechanics} (Butterworth-Heinemann, London, 2005), pp. 2-3.
\bibitem{feynman2} R.P. Feynman, R.B. Leighton, and M.L. Sands, {\it The Feynman Lectures on Physics} (Addison-Wesley, San Francisco, 1989), Vol. II, chapter 19.
\bibitem{moore} T.A. Moore, ``Getting the most out of least action: A proposal", Am. J. Phys. {\bf 72}, 523-527 (2004). 
\bibitem{merzbacher} E. Merzbacher, {\it Quantum Mechanics} (John Wiley and Sons, New York, 1998). 
\bibitem{taylor} C.G. Gray and E.F. Taylor, ``When action is not least", Am. J. Phys. {\bf 75}, 434-458 (2007).
\bibitem{taylor2} J. Hanc, E.F. Taylor, and S. Tuleja, ``Variational mechanics in one and two dimensions", Am. J. Phys. {\bf 73}, 603-610 (2005); 

J. Hanc, E.F. Taylor, and S. Tuleja, ``Deriving Lagrange's equations using elementary calculus'', Am. J. Phys. {\bf 72}, 510-513 (2004);

J. Hanc and E.F. Taylor, ``From conservation of energy to the principle of least action", Am. J. Phys. {\bf 72}, 514-521 (2004). 

E.F. Taylor, ``A call to action'', Am. J. Phys. {\bf 71}, 423-425 (2003).
\bibitem{software} See interactive apps available at http://www.eftaylor.com/leastaction.html. 
\bibitem{bork} The earliest computational attempt, to our knowledge, is by A. Bork and A. Zellweger, ``Least action via Computer", Am. J. Phys. {\bf 37}, 386-390 (1969). 
\bibitem{class} According to the author's experience and informal polling among the colleagues about their student-era experience. 
\bibitem{sage} Sage is an Open Source Mathematics Software available at http://www.sagemath.org.
\bibitem{nr} We limit ourselves to non-relativistic regime.
\bibitem{introphysics} H.D. Young and R.A. Freedman, {\it University Physics} (Addison-Wesley, San Francisco, 2007), chapters 2, 13; D. Halliday, R. Resnick, and J. Walker, {\it Fundamentals of Physics} (Wiley, New York, 2007), chapters 2, 14. 
\bibitem{hibbs} R.P. Feynman and A.R. Hibbs, {\it Quantum Mechanics and Path Integrals} (McGraw-Hill, New York, 1965), pp. 28.
\bibitem{wen} X.-G. Wen, {\it Quantum field theory of many-body systems} (Oxford University Press, New York, 2004) , pp. 24-25.

\end{thebibliography}
\end{document}